\documentclass[sigconf]{acmart}
\usepackage{tikz,pgfplots}
\usepackage{pgfplots}
\usepackage{amsmath}
\usepackage{xcolor}
\usepackage{xspace}
\pgfplotsset{compat=1.18}
\usepackage{cleveref}
\usepackage{algorithm}
\usepackage{algorithmic}
\usepackage{colortbl}
\usepackage[dvipsnames]{xcolor}
\usepackage{makecell} 
\usepackage{multirow}
\usepackage{booktabs}
\usepackage{arydshln}
\definecolor{scienceblue}{RGB}{31, 119, 180}

\AtBeginDocument{%
  }

\copyrightyear{2026}
\acmYear{2026}
\setcopyright{cc}
\setcctype{by}
\acmConference[KDD '26]{Proceedings of the 32nd ACM SIGKDD Conference on Knowledge Discovery and Data Mining V.2}{August 09--13, 2026}{Jeju Island, Republic of Korea}
\acmBooktitle{Proceedings of the 32nd ACM SIGKDD Conference on Knowledge Discovery and Data Mining V.2 (KDD '26), August 09--13, 2026, Jeju Island, Republic of Korea}
\acmDOI{10.1145/3770855.3818326}
\acmISBN{979-8-4007-2259-2/2026/08}

\begin{document}

\title{Breaking the Likelihood Trap: Consistent Generative Recommendation with Graph-structured Model}

\author{Qiya Yang}
\authornote{Equal Contribution.}
\authornote{Work done during an internship at Kuaishou.}
\affiliation{%
  \institution{Peking University}
  \city{Beijing}
  \country{China}
}
\email{kexinqya@gmail.com}

\author{Xiaoxi Liang}
\authornotemark[1]
\affiliation{%
  \institution{Peking University}
  \city{Beijing}
  \country{China}
}
\email{liangxx0419@gmail.com}

\author{Zeping Xiao}
\affiliation{%
  \institution{Kuaishou Technology}
  \city{Beijing}
  \country{China}}
\email{xiaozeping@kuaishou.com}

\author{Ying Cao}
\affiliation{
  \institution{Kuaishou Technology}
  \city{Beijing}
  \country{China}}
\email{caoying03@kuaishou.com}

\author{Yingjie Deng}
\affiliation{%
  \institution{Kuaishou Technology}
  \city{Beijing}
  \country{China}
}
\email{dengyingjie03@kuaishou.com}

\author{Yuxin Ren}
\affiliation{
  \institution{Kuaishou Technology}
  \city{Beijing}
  \country{China}}
\email{renyuxin@kuaishou.com}

\author{Yalong Wang}
\authornote{Corresponding Author.}
\affiliation{
  \institution{Kuaishou Technology}
  \city{Beijing}
  \country{China}}
\email{wangyalong03@kuaishou.com}

\author{Yongqi Liu}
\affiliation{
  \institution{Kuaishou Technology}
  \city{Beijing}
  \country{China}}
\email{liuyongqi@kuaishou.com}

\renewcommand{\shortauthors}{Yang et al.}
\newcommand{\methodname}{\textsc{Congrats}\xspace}
\newcommand{\decodername}{Graph-structured Decoder\xspace}
\newcommand{\graphname}{DAG\xspace}
\newcommand{\modelname}{Graph-structured Model\xspace}
\newcommand{\trainingname}{Consistent Differentiable Training\xspace}
\newcommand{\Lmain}{\LL_{G}}
\newcommand{\LE}{\LL_{G-R}}
\newcommand{\nop}[1]{}

\begin{abstract}
Reranking, as the final stage of recommender systems, plays a crucial role in determining the final exposure, directly influencing user experience.
Recently, generative reranking has gained increasing attention for formulating reranking as a holistic sequence generation task, implicitly modeling complex dependencies among items.
However, most existing methods suffer from the \textit{likelihood trap}, where high-likelihood sequences are often repetitive and perceived as low-quality by humans, thereby limiting user engagement.

In this work, we propose \textbf{Con}sistent \textbf{Gra}ph-\textbf{st}ructured Generative Recommendation (\textbf{\methodname}). We first introduce a novel \modelname, which enables the generation of more diverse sequences by exploring multiple paths. This design not only expands the decoding space to promote diversity, but also improves prediction accuracy by explicitly modeling item dependencies from graph transitions.
Furthermore, we design a \trainingname method that incorporates an evaluator, allowing the model to learn directly from user preferences.
Extensive offline experiments validate the superior performance of \methodname over state-of-the-art reranking methods. 
Moreover, \methodname has been evaluated on a large-scale video-sharing app, Kuaishou, with over 300 million daily active users, demonstrating that our approach significantly improves both recommendation quality and diversity, validating our effectiveness in practical industrial platforms.
\end{abstract}




\maketitle

\section{Introduction}
Recommender systems deliver personalized lists of items to users and are widely deployed across domains such as social media, e-commerce, and video-sharing platforms. 
They impose stringent requirements on real-time inference, accuracy, and diversity. 
In general, these systems follow a multi-stage architecture consisting of matching, ranking, and reranking~\cite{zheng2024full}. 
The final stage, reranking, refines the top-ranked items from the ranking stage and produces a reordered sequence, playing a crucial role in determining the final user exposure and experience.

The core objective of reranking is to explore an optimal sequence that maximizes user feedback~\cite{ren2024non}. 
To achieve this goal, recent reranking methods primarily adopt the Generator–Evaluator framework~\cite{feng2021grn,lin2024discrete,ren2024non,wang2025nlgr,yang2025comprehensive}, where the generator produces multiple candidate item lists, and the evaluator calculates the list-wise utility (e.g., aggregated click-through rate and dwell time) by simulating real user feedback. The list with the highest score is then selected for display.
For the generator, traditional approaches~\cite{pei2019personalized} treat the reranking task as a scoring-and-sorting problem, where the model predicts the utility of each candidate, then returns top-ranked ones. 
However, these paradigms evaluate the utility of each item independently, thereby overlooking the combinatorial dependencies between candidates and often resulting in suboptimality.
Recently, generative approaches~\cite{rajput2023recommender,ren2024non,lin2024discrete,yang2025comprehensive} have emerged.  By formulating reranking as a holistic sequence generation task, these methods inherently capture dependencies, thereby aligning more closely with the reranking objective.
Early generative generators follow an autoregressive (AR) paradigm~\cite{feng2021grn}. Although they achieve promising performance, the step-by-step generation process hinders their deployment in real-time industrial systems. 
Subsequent work has proposed non-autoregressive (NAR) methods~\cite{ren2024non,wang2025nlgr}, achieving real-time inference by eliminating the conditional dependence on output items and generating the entire sequence in one shot solely from the inputs.  

\begin{figure}
    \centering
    \includegraphics[width=0.95\linewidth]{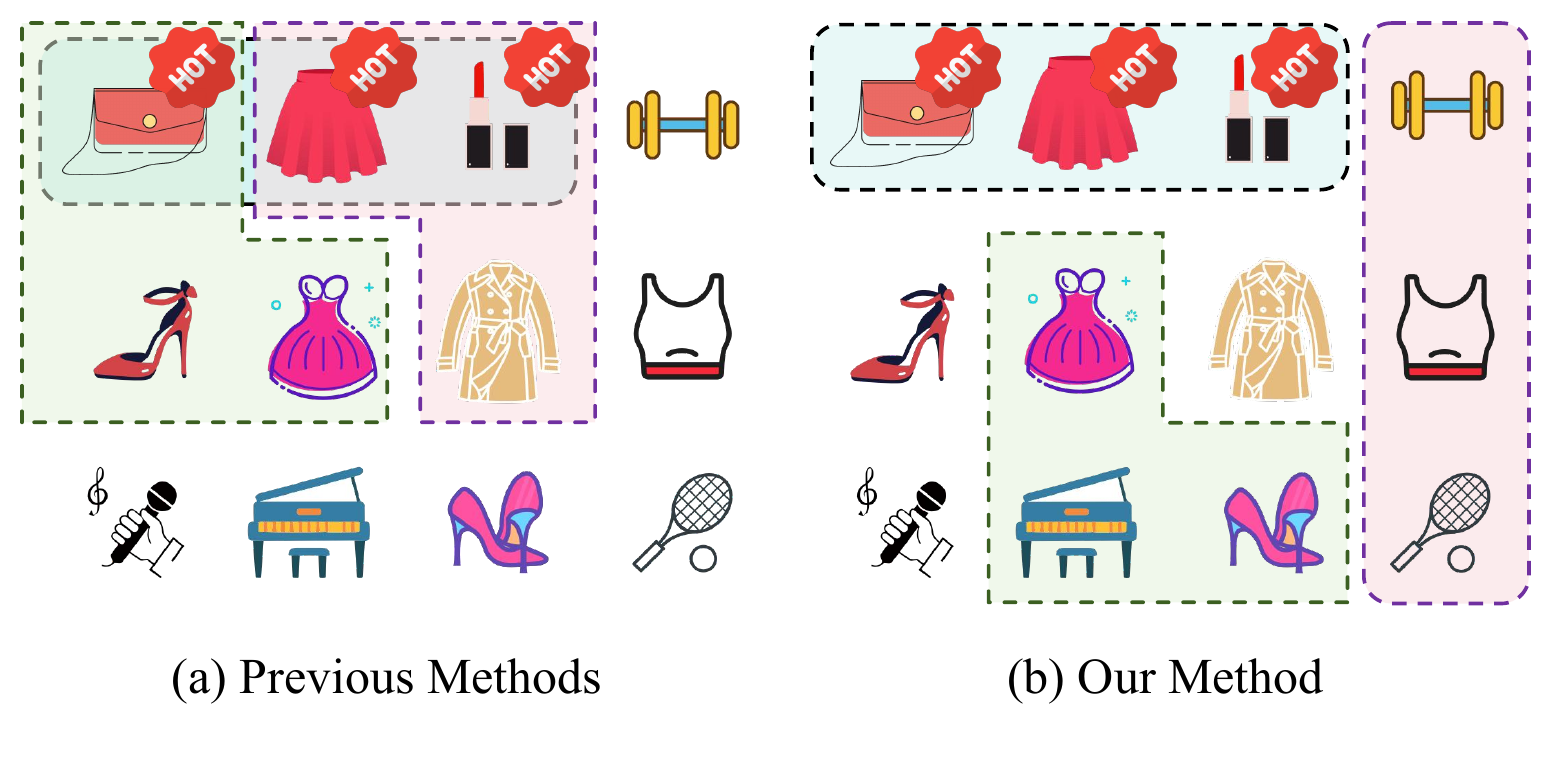}
    \caption{An intuitive understanding of the \textit{likelihood trap}. The dashed box encloses the sequence of three items generated by the same method. Previous generative reranking methods (left) tend to repeatedly recommend high-frequency (``hot'') items, leading to list Homogeneity. 
    In contrast, our proposed \methodname effectively generates diverse sequences (right), thereby achieving more personalized recommendations.
}
\label{fig:diverse}
\end{figure}

Although previous generative methods have achieved notable progress, they still suffer from the severe \textit{likelihood trap}~\cite{zhang2020tradingdiversityqualitynaturallikelihoodtrap,Ott2018AnalyzingUI,welleck2019neuralunlikelihoodtraining}.
Specifically, training the generator to maximize the likelihood of observed samples inherently biases it toward deterministic behavior.
As a consequence, the model repeatedly recommends a small subset of high-frequency items that have already been widely exposed, as illustrated in~\autoref{fig:diverse}.



To this end, we propose a novel \textbf{Con}sistent \textbf{Gra}ph-\textbf{st}ructured Generative Recommendation framework, named \methodname. To guarantee real-time inference, we build upon the architecture of NAR4Rec~\cite{ren2024non} by adopting a non-autoregressive mechanism. 
We first propose a novel \modelname that not only substantially expands the decoding exploration space by structuring hidden states as a directed acyclic graph, unlocking diverse potential outcomes beyond the constraints of a linear chain, 
but also explicitly models item dependencies via a learnable graph transition module to improve prediction accuracy. 
Specifically, we replace the decoder in~\cite{ren2024non} with a directed acyclic graph decoder, extending the set of positional embeddings to constitute the graph vertices.
We introduce the graph transition module with a learnable matrix to explicitly model the connectivity among these vertices, allowing the model to identify the optimal generation path.

\nop{
We design a novel \modelname, that not only substantially expands the decoding exploration space via a graph structure to enhance decoded diversity, but also explicitly models the item dependencies through the graph transition process to improve prediction accuracy.
Specifically, we replace the decoder in~\cite{ren2024non} with a directed acyclic graph decoder, and treat positional embeddings as graph vertices. We set the the number of positional embeddings to several times to expand the decoding space. We then learn a graph transition module to select the final sequence.
}
    
On the other hand, the standard likelihood training method tends to assign too much probability to frequent items, leading to repetitive and dull outputs~\cite{welleck2019neuralunlikelihoodtraining, zhang2020tradingdiversityqualitynaturallikelihoodtrap,holtzmancurious}. 
This is misaligned with user diversity demand.
An intuitive solution is to incorporate user feedback into the training process. 
Hence, we design a \trainingname method that integrates the evaluator into the generator’s training loop, steering the generator toward a consistent direction with user preferences.
Specifically, to achieve direct optimization guided by the evaluator, we leverage the Gumbel-Softmax reparameterization technique~\cite{jang2016categorical}. This relaxes the discrete sampling constraint to make the entire system differentiable, thereby enabling gradients to flow directly from the evaluator back to the generator.

We conduct extensive experiments and demonstrate that \methodname outperforms several state-of-the-art methods.  Additionally, analysis and ablation studies further validate the effectiveness of the proposed tailored designs. The main contributions of our paper are summarized as follows:
\begin{itemize}
\item We first identify the \textit{likelihood trap} in recommender systems and propose a novel framework to address this issue from the perspectives of decoding architecture and user preference alignment.
\item We introduce a novel \modelname that not only enhances the diversity of generated sequences by expanding the decoding space, but also improves accuracy by explicit inter-item dependencies from the graph transitions.
\item We propose \trainingname method, that integrates the evaluator into the training loop, ensuring the generator is optimized for user preference alignment.
\item Extensive offline experiments demonstrate that \methodname outperforms the state-of-the-art methods. We also show significant online improvements in a popular video app, Kuaishou, which serves more than 300 million users daily. 
\end{itemize}

\section{Related Work}
\subsection{Reranking Recommender System}

Reranking has received sustained attention from both academia and industry over the years, as it serves as a critical interface for delivering recommendations to a large user base under stringent requirements of \textit{real-time inference}, \textit{accuracy}, and \textit{diversity}~\cite{covington2016deep}.
Early reranking works often relied on rule-based mechanisms, such as Maximal Marginal Relevance (MMR)~\cite{carbonell1998use} and Determinantal Point Processes (DPP)~\cite{chen2018fast}. These approaches generate the final ranking by optimizing a manually designed objective function that balances accuracy and diversity. Nevertheless, they are fundamentally limited by inflexible rules and an inability to model complex, dynamic user preferences.
Subsequently, Generator–Evaluator (G–E) two-stage methods have emerged~\cite{shi2023pier,ren2024non,wang2025nlgr,yang2025comprehensive}, in which the generator produces a diverse set of candidate sequences and the evaluator assigns list-wise scores to select the optimal one for final exposure.
For the generator, the task is typically formulated as a scoring-and-sorting problem~\cite{pei2019personalized}.
Given an initial ordered list, the model assigns a score to each item and then re-ranks the list in descending order of these scores. By ignoring inter-item dependencies, these methods often yield sub-optimal performance.

Recently, autoregressive (AR) generative models~\cite{feng2021grn} have been adopted to reformulate reranking as a sequence generation task via next-item prediction. Despite their improved performance from implicitly modeling item dependencies, the high computational cost of AR decoding limits their applicability in latency-sensitive industrial settings.
To address this, NAR4Rec\cite{ren2024non} proposed a non-autoregressive (NAR) matching mechanism that calculates the candidate-position matching scores and generates the entire list in parallel.
Although this method achieves minimal latency, it suffers from a limited decoding space, consequently yielding limited performance in terms of recommendation accuracy and diversity.

\nop{
Recommender systems aim to generate personalized item sequences that balance \textit{real-time inference}, \textit{accuracy}, and \textit{diversity}~\cite{covington2016deep}, powering applications ranging from e-commerce and short-video platforms to social media feeds. Reranking is the final stage,  playing a pivotal role in determining the final item exposure. 

Early reranking works often relied on rule-based mechanisms, such as Maximal Marginal Relevance (MMR)~\cite{carbonell1998use} and Determinantal Point Processes (DPP)~\cite{chen2018fast}. They construct the final list by optimizing a hand-crafted objective function that trades off accuracy for diversity. However, these methods are limited by their rigid rules and inability to model complex, dynamic user preferences.

Recently, Generator-Evaluator (G-E) two-stage methods have emerged~\cite{shi2023pier,ren2024non,wang2025nlgr,yang2025comprehensive}, where the generator produces a diverse set of candidate sequences, and the evaluator estimates list-wise scores to select the optimal one for final exposure. 
For generator, conventional approaches typically formulate the task as a scoring-and-sorting problem~\cite{pei2019personalized}: 
Given an initial ordered list, these models calculate ta score for each item and subsequently re-rank the list in descending order of these scores. 
However, such methods often overlooking the combinatorial dependencies and often leading to sub-optimal performance.

Subsequent approaches adopt autoregressive (AR) generative models~\cite{feng2021grn} as the generator, reformulating reranking as a sequence generation task via ``next-item prediction''. 
Distinct from the rigid scoring-and-sorting paradigm, this generative formulation explicitly models high-order item dependencies to optimize list-wise utility.
However, the high computational cost of AR decoding hinders their deployment in latency-sensitive industrial applications.
To address this, \cite{ren2024non} proposed a non-autoregressive (NAR) model that generates the entire list in parallel. Although this method achieves minimal latency, it suffers from a constrained decoding space, consequently yielding limited performance in terms of recommendation accuracy and diversity.
}





\subsection{Likelihood Trap in Generative Models}
The concept of the \textit{likelihood trap} originates from the Natural Language Generation (NLG)~\cite{zhang2020tradingdiversityqualitynaturallikelihoodtrap,welleck2019neuralunlikelihoodtraining,holtzmancurious,chaudhary2022current,meister2023locallyeth}, revealing a counter-intuitive observation in generative models: high probability does not equal high quality~\cite{zhang2020tradingdiversityqualitynaturallikelihoodtrap,meister2023locallyeth}.
This phenomenon arises from the standard Maximum Likelihood Estimation (MLE), which biases the model to assign excessive probability mass to frequent tokens during inference, resulting in dull, repetitive and degenerate outputs~\cite{welleck2019neuralunlikelihoodtraining,holtzmancurious, zhang2020tradingdiversityqualitynaturallikelihoodtrap,chaudhary2022current,meister2023locallyeth}.


To mitigate this, \cite{holtzmancurious} introduced Nucleus Sampling to escape repetition by sampling from the top probability mass, though this introduces stochastic instability.
In response, \cite{su2022contrastive} proposed Contrastive Search, utilizing a deterministic approach to avoid the \textit{likelihood trap}. 
Moreover, modern LLMs typically incorporate Reinforcement Learning from Human Feedback (RLHF)~\cite{achiam2023gpt,brown2020languagegpt3,ouyang2022training,glaese2022improvingrl,knox2009interactivelyrl,Schulman2017ProximalPPO,engstrom2020implementationppotrpo,xu2024dpoppo,zhongdpomeetsppo,rafailov2023directdpo,wang2024unifyingrlhf,Lambert2025ReinforcementLF} as a crucial post-training step. This process slightly refines the probability distribution to better align the model with human preferences.

We argue that a parallel \textit{likelihood trap} persists in generative recommendation.
In this context, the ``dull'' content in NLP corresponds to high-frequency items in recommender systems, manifesting as homogeneous recommendations.
Unlike the vast vocabulary of LLMs, reranking operates on a limited set, resulting in a distinct probability distribution.
Consequently, directly applying LLM sampling strategies is highly prone to selecting tail candidates, leading to a severe loss of relevance, a phenomenon termed \textit{relevance collapse}~\cite{volodkevich2025autoregressive}. 
Additionally, RL-based methods frequently suffer from training inefficiency and instability~\cite{raheja2026rlhfinstability,xie2025dagrpoinstab}.
In light of these limitations, how to escape the \textit{likelihood trap} and ensure diverse and high-utility sequences remains a critical open problem in the recommendation community.


\nop{
The concept of the \textit{likelihood trap} originates from the Natural Language Generation (NLG)~\cite{zhang2020tradingdiversityqualitynaturallikelihoodtrap,welleck2019neuralunlikelihoodtraining,holtzmancurious,chaudhary2022current,meister2023locallyeth}, revealing a counter-intuitive observation in generative models: high probability does not equal high quality~\cite{zhang2020tradingdiversityqualitynaturallikelihoodtrap,meister2023locallyeth}.
This phenomenon arises from the standard Maximum Likelihood Estimation (MLE). Under this objective, the model tends to assign too much probability to sequences containing repeated and frequent words, leading to the generation of dull, repetitive, and degenerate outputs.~\cite{welleck2019neuralunlikelihoodtraining,holtzmancurious, zhang2020tradingdiversityqualitynaturallikelihoodtrap,chaudhary2022current,meister2023locallyeth}.

To mitigate this, \cite{holtzmancurious} introduced Nucleus Sampling to escape repetition by sampling from the top probability mass, though this introduces stochastic instability.
In response, \cite{su2022contrastive} proposed Contrastive Search, utilizing a deterministic approach to avoid the \textit{likelihood trap}. 
\nop{
}
Recently, the emergence of LLMs such as ChatGPT~\cite{achiam2023gpt,brown2020languagegpt3,ouyang2022training} has popularized the Reinforcement Learning (RL)~\cite{Lambert2025ReinforcementLF,ouyang2022training,knox2011augmenting}.
In contrast to MLE's local token optimization,  the Reinforcement Learning (RL)~\cite{achiam2023gpt,brown2020languagegpt3,ouyang2022training} employs Proximal Policy Optimization (PPO)~\cite{Schulman2017ProximalPPO} to target sequence-level global rewards, effectively penalizing high-probability but low-quality artifacts like repetition. Direct Preference Optimization (DPO)~\cite{rafailov2023directdpo} further streamlines this by eliminating the explicit reward model, directly restructuring the probability distribution to suppress \textit{likelihood trap} and align with high-quality text~\cite{silvastabilityworkshop}.

\nop{
%
}

We argue that a parallel \textit{likelihood trap} persists in generative recommendation.
In this context, the ``dull'' content in NLP corresponds to high-frequency items in recommender systems, manifesting as homogeneous recommendations.
Unlike the vast vocabulary of LLMs, reranking operates on a limited candidate set. Directly applying these sampling strategies to generative reranking often yields adverse effects, attributed to the constrained decoding space compared to LLMs. 
Additionally, while RL-based reranking methods like~\cite{feng2021grn} have been proposed, they are frequently plagued by training instability.
Escaping the \textit{likelihood trap} to generate diverse and high-utility sequences remains an open challenge in the recommendation community. 
}


\nop{
\subsection{Non-autoregressive Sequence Generation}
Large language models (LLMs) drive recent advances in natural language processing. 
As their depth and width scale up, they become increasingly computationally and memory-intensive.
To improve efficiency with minimal performance degradation, several techniques have been proposed, such as weight pruning~\cite{fanreducing}, network quantization~\cite{lin2024awq}, matrix decomposition~\cite{wangexploring}, knowledge distillation~\cite{sun2019patient,ren2023tailoring}, and non-autoregressive generation~\cite{gu2018non,gu2019levenshtein}. 

Among these methods, non-autoregressive generation methods generate outputs in parallel to speed up decoding. Traditional autoregressive models must predict output tokens one by one, conditioning each prediction on previously generated outputs. While effective at capturing sequential dependencies, it suffers from slow inference due to its inherently left-to-right decoding and exposure bias, since training uses ground truth tokens while inference relies on model predictions. 
In contrast, non-autoregressive models assume conditional independence among output tokens given the input, allowing the entire sequence to be generated in parallel.  Early approaches, such as the Non-Autoregressive Transformer~\cite{gu2018non} predicts a target length and fertility (number of target tokens per source token), then expand and decode outputs in a single forward pass. Subsequent methods like Mask-Predict~\cite{ghazvininejad2019maskpredict} introduce iterative refinement, where an initial sequence of masked tokens is iteratively completed and improved over multiple decoding steps. These approaches offer significant speedups but often sacrifice output quality.

To narrow the performance gap between non-autoregressive and autoregressive generation, researchers have proposed various enhancements. Latent-variable models, such as Latent Transformer~\cite{kaiser2018fast} and VQ-VAE~\cite{razavi2019generating}, introduce discrete or continuous latent sequences that capture high-level structure, which are then decoded in parallel into final outputs. CTC-based models adapt alignment-free training from speech recognition to text generation~\cite{libovicky2018end}, marginalizing over possible output sequences using blank tokens. 
Despite these advances, non-autoregressive generation still faces challenges in modeling target dependencies, diversifying sequence outputs, and training stability.  
A recent study~\cite{huang2022learning} provides a unified perspective, showing that most existing methods actually modify targets or inputs to reduce token dependencies in the data distribution, which eases training but introduces data distortion.
}

\section{Method}

In this section, we present a detailed introduction of \methodname. 
We first formally define the task in \cref{sec:formulation}. 
Then, we elaborate on the proposed \modelname in \cref{sec:generative model}, highlighting how it leverages graph-structured decoder to enhance recommendation diversity and improve prediction accuracy. 
Finally, we detail the \trainingname method in \cref{sec:consistent}, which is designed to better align model optimization with the overall recommendation objective.

\subsection{Problem Formulation}\label{sec:formulation}
Given \( n \) candidate items from the previous stage, denoted as \( X = \{x_1, x_2, \cdots, x_n\} \), the goal of reranking is to derive an ordered list \( Y = \{y_1, y_2, \cdots, y_m\}\) of length \(m\). In practice, $m$ is usually fewer than ten, while $n$ ranges from several tens to hundreds.
Recent reranking methods typically adopt a generator–evaluator framework~\cite{ren2024non,shi2023pier,yang2025comprehensive}, where the generator produces sequences and the evaluator estimates their expected utility.
Formally, the generator \(\mathcal{G}\) takes the user-specific features \(u\) (including the user profile and interaction history) and candidate items $X$ as input to produce a target list $Y$:
\begin{equation}
Y = \mathcal{G}(u, X; \theta),
\end{equation}
where $\theta$ denotes the learnable parameters of the generator.

Given the sequence $Y$, the evaluator $\mathcal{E}$ predicts the overall utility $R$ by aggregating scores from multiple feedback objectives. This estimation is formulated as:
\begin{equation}
R = \mathcal{E}(u, Y; \phi) = \sum_{k=1}^{K} w_k \hat{r}_k,
\end{equation}
where $\hat{r}_k$ represents the predicted reward signal for the $k$-th feedback type (e.g., views, likes, and clicks) and $w_k$ is the manually assigned weight. Our goal is to optimize the generator's parameters $\theta$ such that the generated list $Y$ maximizes the expected total utility $R$ defined by the evaluator.



\begin{figure*}
    \centering
    \includegraphics[width=0.95\linewidth]{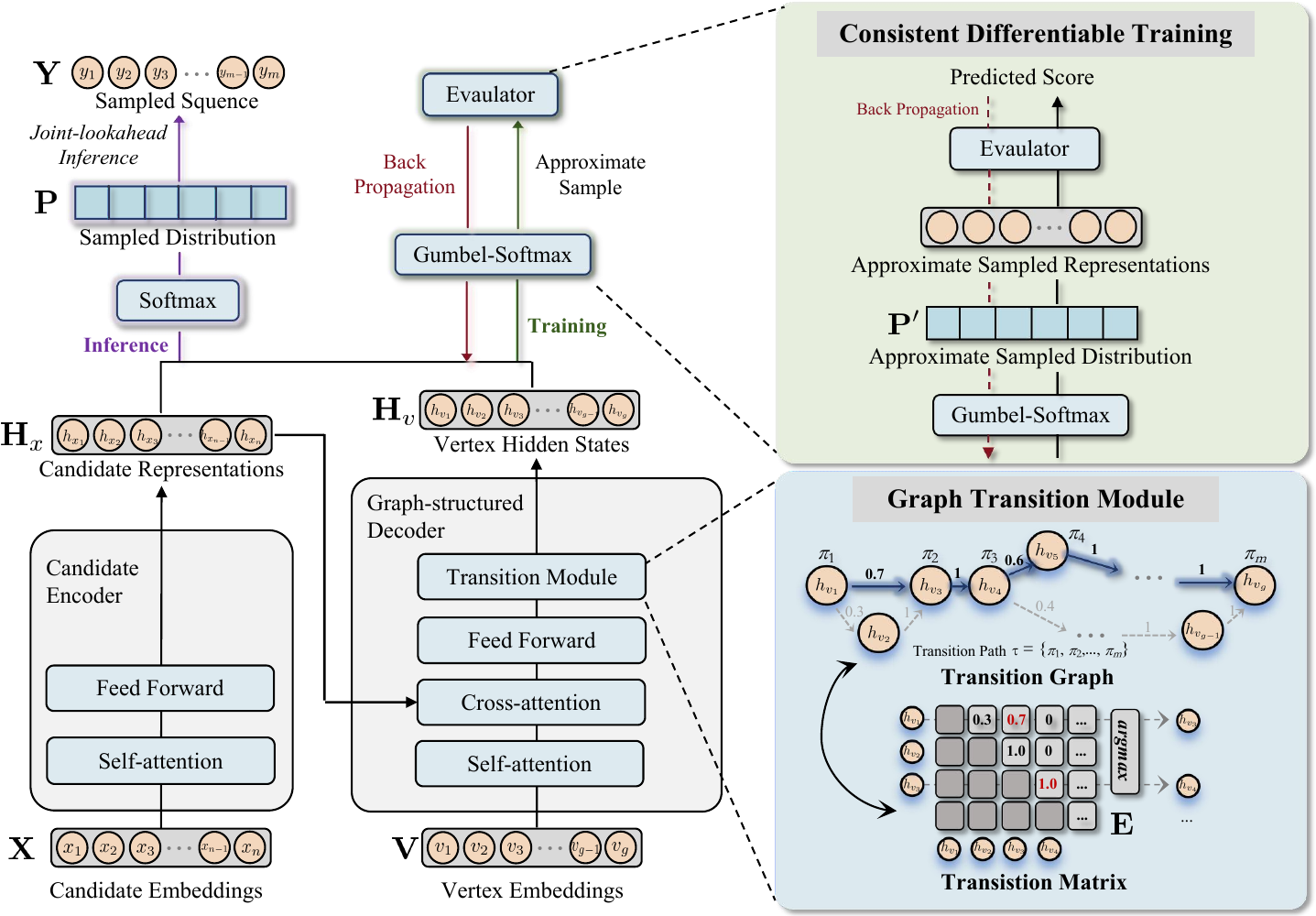}
    \caption{Overview of our proposed \methodname. The model comprises a Candidate Encoder and a Graph-structured Decoder. The decoder incorporates a Graph Transition Module that structures hidden states into a directed acyclic graph to expand the exploration space (Bottom-Right). During training, we introduce the \trainingname method, employing Gumbel-Softmax to approximate the discrete sampling process to allow gradients from the Evaluator to back-propagate directly to the generator (Top-Right). During inference, the model generates sequences via Joint-lookahead decoding (Left). }
    \label{fig:overview}
\end{figure*}

\subsection{Graph-structured Model}\label{sec:generative model}
\nop{
Autoregressive generative methods struggle to apply to the real industry system due to its overhead inference latency.
While the non-autoregressive fashion, generates the entire list at once.

To address these challenges, we designed two key components in our \modelname: a \textbf{candidate encoder} for effectively encoding representations of candidates, and a novel \textbf{graph-structured decoder} to represent the hidden states in a Directed Acyclic Graph (DAG), where each path of the DAG corresponds to a specific recommended sequence. The model generates sequences in an NAR manner while introducing contextual dependency by sampling the graph vertices with small indexes to large indexes. 
}
\nop{
Initially, we randomly initialize an embedding for each vertex in the target sequence. Note that these vertex embeddings are shared across training data to enhance learning on sparse data. Subsequently, bidirectional self-attention and cross-attention modules are incorporated to acquire representations for each vertex, leveraging information from the users and candidates from the encoder.

At last, instead of applying $\text{Softmax}$ over a fixed vocabulary as in text generation~\cite{achiam2023gpt},  we perform $\text{Softmax}$ over the changing candidates, allowing the model to learn the item distributions at each vertex. Each candidate is matched with every vertex in the target sequence. In the following, we provide a detailed description of the \genmodelname.
}

As illustrated in~\autoref{fig:overview},  we adopt the encoder-decoder architecture and the matching mechanism following~\cite{ren2024non}.  Although the non-autoregressive paradigm guarantees real-time inference, its small and deterministic decoding space tends to produce repetitive and similar sequences, as it consistently selects the highest-probability item at each position.
In this way, we expand the number of graph vertices \( g \) to be several times the original length of positions embeddings \( m \). 
This design structures the decoder's hidden states as a directed acyclic graph rather than a traditional linear chain, effectively enlarging the decoding space to enable richer sampling pathways for enhanced sequence diversity. Furthermore, by employing learnable graph transitions module to explicitly model item dependencies, we significantly improve prediction accuracy. We detail the proposed workflow below.

\noindent
\textbf{Candidate Encoder }  Given \( n \) candidate items \( X = \{x_1, \cdots, x_n\} \), we first concatenate each item \( x_i \) with the user features \( u \) to form the input matrix \( \mathbf{X} \in \mathbb{R}^{n \times d} \).
This matrix is processed by a standard Transformer encoder~\cite{vaswani2017attention} consisting of $L$ stacked blocks, where each block comprises a self-attention layer and a feed-forward network.
Consequently, we obtain the final encoded representations as:
\begin{equation}
    \mathbf{H}_x =  [h_{x_1}, \cdots, h_{x_n}] = \text{Encoder}(\mathbf{X}), 
\end{equation}
where $\mathbf{H}_x$ represents the learned candidate embeddings.

\noindent
\textbf{Graph-structured Decoder }
The graph-structured decoder comprises $ L $ stacked blocks, each consisting of a self-attention layer, a bidirectional cross-attention layer, and a feed-forward network.
We initialize the decoder input as a set of learnable embeddings $\mathbf{V} = [v_1, \cdots, v_g]$, referred to as position vertices.
These vertex embeddings maintain the same dimension as the candidate features and utilize an expanded sequence length of $g = \lambda m$, where $\lambda$ serves as a scaling factor.
Specifically, the embeddings $\mathbf{V}$ are updated by the self-attention layer and subsequently query the candidate hidden states $\mathbf{H}_x$ via the cross-attention mechanism:
\begin{gather}
    \mathbf{H}_v = \text{Softmax}\left(\frac{\mathbf{Q}_c \mathbf{K}_c^\intercal}{\sqrt{d}}\right)\mathbf{V}_c, \label{eq:cross_attn} \\
    \mathbf{Q}_c = \mathbf{V} \mathbf{W}_c^\text{Q}, \quad
    \mathbf{K}_c = \mathbf{H}_x \mathbf{W}_c^\text{K}, \quad
    \mathbf{V}_c = \mathbf{H}_x \mathbf{W}_c^\text{V}, \notag
\end{gather}
where $\mathbf{W}_c^\text{Q}$, $\mathbf{W}_c^\text{K}$, and $\mathbf{W}_c^\text{V}$ are linear projection matrices and $d$ represents the embedding dimension.
Formally, the overall process is formulated as:
\begin{equation}
    \mathbf{H}_v = [h_{v_1}, \cdots, h_{v_g}] = \text{Decoder}(\mathbf{V}),
\end{equation}
where $\mathbf{H}_v$ denotes the generated vertex hidden states.

\nop{
\begin{equation}
\label{eq:cat}
\text{Cross-Attention}(\mathbf{Q}_v, \mathbf{K}_{H_c}^T, \mathbf{H}_x)
= \text{Softmax}\left(\frac{\mathbf{Q}_v \mathbf{K}_{H_c}^T}{\sqrt{d}}\right)\mathbf{H}_x,
\end{equation}
where $\mathbf{Q}_v$ is the linear projection of $\mathbf{V}$, $\mathbf{K}_{H_c}^T$ is the linear projection of $\mathbf{H}_x$, and $d$ represents the embedding dimension. 
}

\noindent
\textbf{Graph Transition }  
Subsequently, we construct a graph transition matrix to sample the final sequence of $m$ vertices. We derive the transition matrix $\mathbf{E} \in \mathbb{R}^{g \times g}$ from the vertex hidden states $\mathbf{H}_v$ via a self-attention-like mechanism:
\begin{gather}
\mathbf{E} = \text{Softmax}\left(\frac{\mathbf{Q}_e \mathbf{K}_e^\intercal}{\sqrt{d}}\right), \label{eq:matrixE}   \ \ \
\mathbf{Q}_e = \mathbf{H}_v \mathbf{W}_e^\text{Q},\quad
\mathbf{K}_e = \mathbf{H}_v \mathbf{W}_e^\text{K},
\end{gather}
where $\mathbf{W}_e^\text{Q}$ and $\mathbf{W}_e^\text{K}$ are learnable projection parameters.
We apply an upper-triangular mask to $\mathbf{E}$ to strictly enforce sequential transitions (i.e., prohibiting $u \to v$ where $u \ge v$).  Subsequently, we perform sequential argmax operations on the masked transition matrix to derive a vertex path $\tau = \{ \pi_1, \pi_2, \dots, \pi_m \}$, where each vertex $\pi_t$ corresponds to the $t$-th output position (as shown on the bottom-right of \autoref{fig:overview}, the chosen vertex path $\tau$ is marked by a solid arrow).
The probability of $\tau$ is formulated as the product of transition probabilities along the vertex path:
\begin{gather}
    P_{\theta}(\tau|X) 
    = \prod_{t=1}^{m-1} P_{\theta}(\pi_{t+1}|\pi_{t}, X) 
    = \prod_{t=1}^{m-1} \mathbf{E}_{\pi_{t}, \pi_{t+1}}, 
\end{gather}
subject to the vertex index constraint $1 = \pi_1 < \pi_2 < \cdots < \pi_m = g$.
Here, $\mathbf{E}_{\pi_{t}, \pi_{t+1}}$ denotes the transition probability from vertex $\pi_t$ to $\pi_{t+1}$. Notably, the computation of the transition matrix is fully parallelizable, ensuring efficiency.

 \noindent
\textbf{Item Prediction  }  
In addition, we learn an item prediction matrix to estimate the probability distribution over candidate items along the vertex path. 
Specifically, we follow the matching mechanism~\cite{ren2024non} and perform a column-wise Softmax operation conditioned on the candidate representations $\mathbf{H}_x$ and vertex states $\mathbf{H}_v$ to derive the  item prediction matrix $\mathbf{P} \in \mathbb{R}^{n \times g}$:
\begin{equation}
    \mathbf{P} = \text{Softmax}(\mathbf{H}_x \mathbf{H}_v^\intercal), \label{eq:matrixP}
\end{equation}
where $\mathbf{P}_{i,j}$ denotes the probability of selecting the $i$-th candidate item at the $j$-th vertex.
Combining the graph transition and item prediction probabilities, we formulate the probability of generating the target list $Y$ as:
\begin{equation}
    P_{\theta}(Y, \tau|X) = \underbrace{\prod_{t=1}^{m-1} \mathbf{E}_{\pi_t, \pi_{t+1}}}_{\text{Transition Probability}} \cdot \underbrace{\prod_{t=1}^{m} \mathbf{P}_{y_t, \pi_t}}_{\text{Prediction Probability}},
\end{equation}
where $\mathbf{E}_{\pi_t, \pi_{t+1}}$ is the transition probability from vertex $\pi_t$ to $\pi_{t+1}$ (defined in~\autoref{eq:matrixE}), and $\mathbf{P}_{y_t, \pi_t}$ is the prediction probability of the target item $y_t$ at vertex $\pi_t$ (defined in ~\autoref{eq:matrixP}).

 \noindent
\textbf{Objective  }  
Since the specific path $A$ is unobserved, we optimize the objective by marginalizing the probability over all possible valid paths $\Gamma$. The generative loss is defined as:
\begin{equation}
    \mathcal{L}_{gen} = - \log P_{\theta}(Y | X) = - \log \sum_{\tau \in \Gamma} P_{\theta}(Y, \tau | X), \label{eq:Lg}
\end{equation}
This formulation encourages the model to discover the most plausible transitions and item selections that reconstruct the ground-truth sequence. Note that the summation over all paths $\Gamma$ can be efficiently computed via dynamic programming algorithms during training~\cite{graves2006connectionistdynamic}.

\nop{
\label{sec:inference}
\noindent
\textbf{Inference }
During inference, the generator first samples a path through the transition matrix \(\mathbf{E}\) and then selects the target candidates based on the output distribution matrix \(\mathbf{P}_g\) along the chosen path.
We adopt a \textit{lookahead search} strategy~\cite{huang2022directed}, jointly considering graph transitions and candidate predictions, rather than making independent local decisions as in \textit{greedy decoding}. 
This joint formulation enables the model to anticipate upcoming steps and make globally consistent choices.
Formally, the joint probability $P_{\theta}(Y, A|X)$ can be factorized as:
\begin{align}
    P_{\theta}(y_1|a_1, X)
    \prod_{i=2}^{m}
    P_{\theta}(a_{i}|a_{i-1}, X)
    P_{\theta}(y_{i}|a_{i}, X),
    \label{eq:sequence_decode}
\end{align}
where each step involves both the transition from $a_{i-1}$ to $a_i$ and the generation of item $y_i$. 
We thus determine the next vertex and its corresponding item jointly by maximizing their combined probability:
\begin{align}
    a^*_{i}, y^*_{i} 
    = \arg\max_{a_i, y_i} 
    P_{\theta}(a_{i}|a_{i-1}, X)
    P_{\theta}(y_{i}|a_{i}, X).
    \label{eq:joint}
\end{align}

This lookahead decoding preserves a diverse decoding space while effectively integrating the structural dependencies captured by the transition sampling, leading to more accurate generation. All operations are executed in parallel as illustrated in~\cref{algo:greedy}, resulting in negligible computational overhead compared with existing methods.
\begin{algorithm}[!t]
   \caption{Lookahead Search Pseudo-code}
   \label{algo:greedy}
\begin{algorithmic}
   \STATE {\bfseries Input:} Graph size $g$, transition matrix $\mathbf{E} \in \R^{g \times g}$, candidate distributions $\mathbf{P}_g \in \R^{n \times g}$
   \STATE \# Jointly consider transitions and candidate likelihoods
   \STATE $\mathbf{E} := \mathbf{E} \otimes [\mathbf{P}_g$.\textsc{max}({dim}=1).\textsc{unsqueeze}({dim}=0)$]$
   \STATE \# $\otimes$ denotes element-wise multiplication
   \REPEAT
       \STATE $i := $ \textit{edges}[$i$] \quad \text{\# move along the selected transition}
       \STATE \textit{output}.\textsc{append}(\textit{tokens}[$i$])
   \UNTIL{$i = m$}
\end{algorithmic}
\end{algorithm}
}

\noindent
\textbf{Inference   }\label{sec:inference}
While the training objective involves marginalizing over all valid paths to optimize the model parameters, the goal of inference is to identify the most probable sequence.
Instead of making independent local decisions (i.e., selecting a path first and then items), we adopt a \textit{joint lookahead search} strategy~\cite{huang2022directed}. This approach simultaneously evaluates the transition probability and the prediction likelihood to determine the optimal transitions and items.

Given the current vertex $\pi_{t-1}$, we determine the next vertex $\pi_t$ and the corresponding target item $y_t$ by maximizing their joint probability product:
\begin{align}
    \pi^*_{t}, y^*_{t} 
    = \arg\max_{\pi_t, y_t} \left( 
    \underbrace{P_{\theta}(\pi_{t}|\pi_{t-1}, X)}_{\text{Transition}} 
    \cdot 
    \underbrace{P_{\theta}(y_{t}|\pi_{t}, X)}_{\text{Prediction}}
    \right),
    \label{eq:joint_inference}
\end{align}
where the first term comes from the transition matrix $\mathbf{E}$ and the second from the prediction matrix $\mathbf{P}$.
By incorporating the prediction score into the transition selection, the model "looks ahead" to ensure that the selected vertex is capable of generating a high-confidence item.
Specifically, we pre-compute a weighted transition matrix based on maximum item probabilities as illustrated in Algorithm~\autoref{algo:inference}. This allows for computationally efficient inference with negligible overhead compared to previous approaches.
\begin{algorithm}[!t]
    \caption{Joint Lookahead Inference}
    \label{algo:inference}
    \begin{algorithmic}[1]
    \REQUIRE Transition matrix $\mathbf{E} \in \mathbb{R}^{g \times g}$, Prediction matrix $\mathbf{P} \in \mathbb{R}^{n \times g}$, Target Length $m$
    \ENSURE Output sequence $Y$
    
    \STATE $s_u \leftarrow \max_{i} \mathbf{P}_{i, u} \quad \forall u \in \{1, \dots, g\}$ \quad \textit{// Pre-compute best item probability for each vertex}
    
    \STATE $\tilde{\mathbf{E}} \leftarrow \mathbf{E} \cdot \text{diag}(\mathbf{s})$ 
    
    \STATE $\pi_0 \leftarrow \text{start\_node}; \ Y \leftarrow \emptyset$
    \FOR{$t = 1$ \TO $m$}
        \STATE $\pi_t \leftarrow \arg\max_{v} \tilde{\mathbf{E}}_{\pi_{t-1}, v}$ \quad \textit{// Select the best next vertex}
        
        \STATE $y_t \leftarrow \arg\max_{i} \mathbf{P}_{i, \pi_t}$   \quad \textit{// Select best item at vertex $\pi_t$}
        
        \STATE $Y \leftarrow Y \cup \{y_t\}$
    \ENDFOR
    
    \RETURN $Y$
    \end{algorithmic}
\end{algorithm}

\nop{
\begin{algorithm}[!t]
   \caption{Lookahead Search Pseudo-code}
   \label{algo:greedy}
\begin{algorithmic}
   \STATE {\bfseries Input:} Graph size $g$, transition matrix $\mathbf{E}$, prediction matrix $\mathbf{P} $
   \STATE \# Jointly consider transitions and prediction likelihoods
   \STATE $\mathbf{E} := \mathbf{E} \otimes [\mathbf{P}$.\textsc{max}({dim}=1).\textsc{unsqueeze}({dim}=0)$]$ \# $\otimes$ denotes element-wise multiplication
   \REPEAT
       \STATE $i := $ \textit{edges}[$i$] \quad \text{\# move along the selected transition}
       \STATE \textit{output}.\textsc{append}(\textit{tokens}[$i$])
   \UNTIL{$i = m$}
\end{algorithmic}
\end{algorithm}
}

\subsection{Consistent Differentiable Learning}
\label{sec:consistent}
\cite{holtzmancurious,zhang2020tradingdiversityqualitynaturallikelihoodtrap,welleck2019neuralunlikelihoodtraining, chaudhary2022current,meister2023locallyeth} demonstrates that a high likelihood does not necessarily imply high human-perceived quality in machine translation.
This discrepancy is particularly critical in recommender systems, where the ultimate goal is to maximize user feedback and satisfy personalized needs.
To bridge this gap, we incorporate the evaluator into the generator’s training loop to explicitly optimize the generator for better alignment with personalized user preferences.

\noindent
\textbf{Training of Evaluator  }
In practice, our online evaluator is instantiated with the Progressive Layered Extraction (PLE) architecture~\cite{10.1145/3383313.3412236}, incorporating multiple task-specific towers to model distinct user behaviors.  
To ensure it functions as a reliable judge, the evaluator is supervised by the cross-entropy loss using the real data:
\begin{equation}
\mathcal{L}_{eval} = -\sum_{k=1}^{K} \sum_{s=1}^N \left[ y_{s,k} \log(\widehat{y}_{s,k}) + (1 - y_{s,k}) \log(1 - \widehat{y}_{s,k}) \right], \label{eq:eval loss}
\end{equation}
where $y_{s,k}$ denotes the ground-truth label for the $k$-th task on the $s$-th sample, $\widehat{y}_{s,k}$ represents the predicted probability, $N$ denotes the number of training samples, and $K$ is the number of target objectives.

\noindent
\textbf{Consistent Differentiable Training method} 
To incorporate the evaluator’s guidance, we sample lists from the generator and feed them into the evaluator to obtain the corresponding utility values \(\widehat{y}_{i,j}\), subsequently backpropagating gradients to steer the optimization direction.
However, the discrete sampling operation is non-differentiable, which blocks gradient back-propagation.
To address this, we apply the Gumbel-Softmax reparameterization~\cite{jang2016categorical,liu2021neural,huijben2022review} to derive a differentiable approximation:
\nop{
}
\begin{equation} \label{eq:gmbstm}
\mathbf{{P'}} = \mathrm{Softmax}\!\left(\frac{\mathbf{H}_x\mathbf{H}_v^\intercal + \mathbf{r}}{T}\right), \quad \mathbf{r} \sim \mathrm{Gumbel}(0,1),
\end{equation}
where \(\mathbf{H}_x\mathbf{H}_v^\intercal\) represents the unnormalized logits, \(\mathbf{r}\) is the Gumbel noise, and \(T\) is a temperature hyperparameter.
Lowering \(T\) sharpens the distribution towards a one-hot vector, effectively approximating hard sampling while remaining differentiable.

To encourage the generator to produce high-utility candidates, we assign a target pseudo-label $y=1$ (indicating positive feedback) to all generated samples. Consequently, the standard cross-entropy objective simplifies to maximizing the log-likelihood of the evaluator's scores:
\begin{equation}
    \mathcal{L}_{con} = - \sum_{k=1}^{K} \sum_{s=1}^{N} \log(\widehat{y}_{s,k}), 
    \label{eq:con loss}
\end{equation}
where $\widehat{y}_{s,k}$ denotes the predicted score of the evaluator for the $s$-th generated sample on the $k$-th task.

In practice, we combine the original generative loss in~\autoref{eq:Lg} to form the final total objective:
\begin{equation} \label{eq:losstotal}
    \mathcal{L}_{total} = \mathcal{L}_{con} + \alpha \mathcal{L}_{gen},
\end{equation}
with $\alpha$ serving as a hyperparameter to balance the consistent alignment loss and the generative loss.
By combining these two objectives, our \methodname produces high-quality sequences that satisfy the evaluator's criteria and achieve superior performance in online tests, as presented in the next section.

\section{Experiments}
In this section, we conduct extensive offline experiments and online A/B tests to demonstrate the performance and efficiency of our proposed \methodname. We first describe the datasets and baselines in~\cref{sec:experiment setup}. Then, offline experiments are presented in~\cref{sec:offline}. The online A/B test results are shown in ~\cref{sec:online exp}. Further ablation and analysis studies are provided in~\cref{sec: analysis} and~\cref{sec:ablation}.

\subsection{Experiment Setup}\label{sec:experiment setup}
\noindent
\textbf{Datasets } 
To validate the superior performance of our proposed \methodname, we conduct extensive experiments in both offline and online settings. For offline evaluation, we utilize the public Avito dataset and a static historical log collected from the Kuaishou video-sharing platform. 
The detailed introduction is as follows:
\begin{itemize}
    \item \textbf{Kuaishou}:  A large-scale dataset from the Kuaishou short-video platform. Each sample represents a real user request with user attributes (e.g., ID, age, gender), 60 candidate videos, and 6 exposed items with interaction feedback.
    \item \textbf{Avito}: A publicly available dataset of user search logs from \textit{avito.ru}. Each record corresponds to a search page containing multiple ads. We use the first 21 days of logs for training and the last 7 days for testing.
\end{itemize}

More statistics are presented in~\autoref{table:dataset_overview}.

\begin{table}[ht]
\caption{Statistics of datasets used in offline experiments.}
\centering
\begin{tabular}{c|cccc}
\toprule
 & \textbf{\#Requests} & \textbf{\#Users} & \textbf{\#Items/Ads} &\textbf{Seq. Len.} \\
\midrule
Kuaishou & 230.5M & 3.2M & 98.5M &6 \\
Avito & 53.6M & 1.3M & 23.6M &5 \\
\bottomrule
\end{tabular}
\label{table:dataset_overview}
\end{table}

\noindent
\textbf{Baselines } For offline experiments, we compare with state-of-the-art reranking methods, including the DNN~\cite{covington2016deep}, DCN~\cite{wang2017deep}, and PRM~\cite{pei2019personalized}, as one-stage generator-only methods.
Edge-Rerank~\cite{gong2022real}, PIER~\cite{shi2023pier}, Seq2slate~\cite{bello2018seq2slate}, NAR4Rec~\cite{ren2024non} as two-stage approaches. 
For online A/B tests, the baseline is NAR4Rec~\cite{ren2024non}. The Ensemble Sort~\cite{Sener2018MultiTaskLAensemblesort} is a baseline in further analysis study.
We briefly introduce these baselines as:
\begin{itemize}
\item  \textbf{Ensemble Sort}~\cite{Sener2018MultiTaskLAensemblesort}: A rule-based strategy that fuses multi-task ranking scores via hybrid multiplicative or additive scoring formulas with manually assigned weights.

    \item \textbf{DNN}~\cite{covington2016deep}: A fundamental click-through rate (CTR) prediction model that employs a standard multi-layer perceptron (MLP) to learn nonlinear feature interactions.
    
    \item \textbf{DCN}~\cite{wang2017deep}: Extends DNN by introducing explicit feature crossing at each layer, thereby capturing higher-order feature interactions without manual feature engineering.
    
    \item \textbf{PRM}~\cite{pei2019personalized}: Models mutual dependencies among items using a self-attention mechanism and ranks items according to their predicted relevance scores.
    
    \item \textbf{Edge-Rerank}~\cite{gong2022real}: Generates context-aware recommendation sequences by performing adaptive beam search over estimated item scores.
    
    \item \textbf{PIER}~\cite{shi2023pier}: Utilizes a hashing-based retrieval strategy to efficiently select top-$k$ candidates from the full permutation space.
    
    \item \textbf{Seq2Slate}~\cite{bello2018seq2slate}: An autoregressive sequence-to-sequence model built upon pointer networks, which iteratively predicts the next item conditioned on previously selected ones.
    
    \item \textbf{NAR4Rec}~\cite{ren2024non}: A non-autoregressive generative recommender that leverages a matching mechanism to generate all items in parallel.
\end{itemize}

\noindent
\textbf{Metrics } 
For offline evaluation, we follow prior works~\cite{shi2023pier,ren2024non} and adopt commonly used metrics on all datasets.  
For the Kuaishou dataset, where $n = 60$ and $m = 6$, we report Recall@6 and Recall@10, measuring the accuracy of selecting the top-$6$ exposed items among 60 candidates.
For the Avito dataset, where the input and output lengths are equal ($n = m = 5$), we evaluate models using AUC and NDCG, focusing on item-wise click-through rate prediction within each list.  
For online evaluation, we rely on the Kuaishou A/B testing platform and monitor a set of engagement metrics, including \textit{Views} (the number of video watched), \textit{Effective Views} (the number of valid views exceeding a duration threshold), \textit{Long Views} (the number of videos with substantial watch duration), \textit{Complete Views} (the number of videos watched to completion), \textit{Likes} (the number of likes), and \textit{Watch Time} (the total duration of video consumption). Among these, \textit{Views} and \textit{Watch Time} are considered the key metrics.

\noindent
\textbf{Implementation } 
All experiments are implemented in TensorFlow. We utilize the Adam optimizer with a learning rate of $10^{-3}$. The batch size is set to 1,024 for the Kuaishou dataset and 256 for the Avito dataset, while other optimization settings remain consistent. Regarding the model hyperparameters, we set the stack depth of the model $L=3$, the graph size factor $\lambda=4$ (corresponding to 24 vertices).
For the Gumbel-Softmax temperature, we conduct experiments over a range of $T \in [0.1, 1.0]$ and observe that \methodname is not sensitive to this hyperparameter. Therefore, we set $T = 0.3$ in all experiments.
For Consistent Differentiable Training method, the number of target objectives $K$ is 3 in~\autoref{eq:eval loss} and~\autoref{eq:con loss}, including \textit{shows}, \textit{clicks}, and \textit{next-slides}. The total loss balancing weight $\alpha$ is set to 0.5 in~\autoref{eq:losstotal}.

\subsection{Offline Experiments} \label{sec:offline}
\autoref{tab:industry_dataset} and ~\autoref{tab:avito_dataset} present the offline results on two datasets. We observed that our \methodname outperforms all baselines, including the state-of-the-art methods.
%
Specifically, our proposed \methodname exhibits significant improvement gain, achieving a 7\% increase in Recall@6 compared to the baseline model NAR4Rec~\cite{ren2024non}.
The superiority mainly stems from the specific design of our \modelname.
Unlike conventional non-autoregressive models that often struggle to capture complex relationships~\cite{ren2024non}, our method organizes the decoder's hidden representations into a graph. This structure allows conflicting items to be assigned to distinct vertices, inherently mitigating the risk of erroneous predictions. Furthermore, by explicitly modeling item dependencies through the graph transition module, we achieve substantial improvements in overall accuracy.

\begin{table}[ht]
    \centering
    \caption{Offline comparison between \methodname and baseline methods for the Kuaishou dataset. 
    The dashed line separates traditional methods (above) from generative methods (below).
    The best results are highlighted in \textbf{bold}, and the second-best results are \underline{underlined}.}
    \begin{tabular}{l|cc}
    \toprule
    Method & Recall@6 & Recall@10 \\
    \midrule
    DNN & 59.47\% & 65.65\% \\
    DCN & 60.22\% & 67.95\% \\
    PRM & 63.17\% & 72.25\% \\
    Edge-rerank & 63.63\% & 71.90\% \\
    PIER & 64.95\% & 72.44\% \\
    \noalign{\vskip\aboverulesep}
    \hdashline
    \noalign{\vskip\belowrulesep}  
    NAR4Rec & \underline{65.05\%} & \underline{73.16\%} \\
    \textbf{\methodname (Ours)} & \textbf{72.84\%} & \textbf{81.83\%} \\
    \bottomrule
    \end{tabular}
    \label{tab:industry_dataset}
\end{table}

\begin{table}[ht]
    \centering
    \caption{Offline comparison between \methodname and baseline methods for the Avito dataset. 
    The dashed line separates traditional methods (above) from generative methods (below).
    The best results are highlighted in \textbf{bold}, and the second-best results are \underline{underlined}.}
    \begin{tabular}{l|cc}
    \toprule
     Method & AUC & NDCG \\
    \midrule
        DNN & 0.6614 & 0.6920 \\
        DCN & 0.6623 & 0.7004 \\
        PRM & 0.6881 & 0.7380 \\
        Edge-rerank & 0.6953 & 0.7203 \\
        PIER & 0.7109 & 0.7401 \\
  \noalign{\vskip\aboverulesep}
    \hdashline
    \noalign{\vskip\belowrulesep}  
        Seq2Slate & 0.7134 & 0.7225 \\
        NAR4Rec & \underline{0.7234} & \underline{0.7409} \\
        \textbf{\methodname (Ours)} & \textbf{0.7541} & \textbf{0.7553} \\
    \bottomrule
    \end{tabular}
    \label{tab:avito_dataset}
\end{table}

\begin{table}[ht]
    \centering
    \caption{
    Comparison of training and inference time between \methodname and baseline models for the Avito dataset. The dashed line separates traditional methods (above) from generative methods (below).
    All experiments are conducted on a Tesla T4 GPU (16G) with a batch size of 1024. 
    The reported times are averaged over 100 iterations.
    }
    \begin{tabular}{l|cc}
    \toprule
      Method   & Training Time (s) & Inference Time (s) \\
    \midrule
    PRM & 0.129 & 0.046 \\
    Edge-rerank & 0.112 & 0.041 \\
    PIER & 0.173 & 0.073 \\ 
      \noalign{\vskip\aboverulesep}
    \hdashline
    \noalign{\vskip\belowrulesep} 
    Seq2Slate & 0.565 & 0.194 \\
    NAR4Rec & \textbf{0.122} & \textbf{0.043} \\ 
    \textbf{\methodname (Ours)} & \underline{0.125} & \underline{0.045} \\
    \bottomrule
    \end{tabular}
    \label{tab:sample_time}
\end{table}

We further compare the training and inference efficiency of \methodname with baseline models on the Avito dataset, as shown in~\autoref{tab:sample_time}.
Among these generative methods, the non-autoregressive model NAR4Rec naturally achieves faster training and inference compared to the autoregressive approach Seq2Slate, as it generates all items simultaneously rather than sequentially. 
Similarly, our proposed \methodname attains comparable efficiency to NAR4Rec, benefiting from its largely parallel calculating mechanism.



\subsection{Online Experiments}
\label{sec:online exp}

We conduct extensive A/B tests on the Kuaishou industrial platform to evaluate the real-world effectiveness of \methodname. 
The primary online test is deployed for a continuous period of five days, involving 5\% of the overall traffic, which ensures a stable and large-scale evaluation setting. 
The online baseline is NAR4Rec~\cite{ren2024non}, a strong non-autoregressive generative reranking model deployed in the production system. 


As shown in~\autoref{tab:online result}, \methodname outperforms the online baseline by \textit{a large margin} (note that improvements greater than 0.2\% in \textit{Views} and 0.5\% in \textit{Likes} are considered highly significant on the Kuaishou platform). These results indicate increased video consumption (evidenced by higher \textit{Views},\textit{ Effective Views}, \textit{Long Views}, and \textit{Complete Views}) and enhanced user satisfaction (reflected by the improvement in \textit{Likes}).
\begin{table}[ht]
\setlength{\tabcolsep}{3pt}
\caption{Online A/B test results. All values indicate relative improvements over the online baseline NAR4Rec. }
\begin{tabular}{ccccc}
\toprule
Views & Effective Views  & Long Views & Complete Views  &Likes\\
\midrule
+0.780\%& +1.301\%& +2.180\% & +3.016\%&+0.515\%\\
\bottomrule
\end{tabular}
\label{tab:online result}
\end{table}


\subsection{Analysis Study} \label{sec: analysis}


We conduct a series of analyses to investigate the detailed characteristics of \methodname, specifically focusing on the cross-list diversity of generated results and the effect of the graph size factor. 
The former evaluates whether \methodname can produce less repetitive and more personalized recommendation lists, while the latter studies how enlarging the graph-structured decoding space affects both recommendation accuracy and online latency. 

\noindent
\textbf{Cross-list Diversity } 
To evaluate the diversity among the generated lists, we employ the pairwise Jaccard similarity~\cite{bag2019efficientjaccard} as the \textit{diversity score}.  Given a set of generated lists $S = \{S_1, \dots, S_n\}$, the score is defined as:
\begin{align}
    \mathrm{Diversity\ Score} = 1 - \frac{2}{n(n-1)} \sum_{i<j} \frac{|S_i \cap S_j|}{|S_i \cup S_j|},
\end{align}
A higher score indicates less overlap across lists, signifying greater diversity.
Additionally, we report the \textit{Repetition Rate}, which measures the average fraction of shared items between list pairs (lower is better).
We also evaluate \textit{Item Coverage}, which quantifies the number of distinct items from the entire candidate pool recommended at least once.
Finally, \textit{Distinct-2} captures the local diversity within each list. It is defined as the ratio of unique consecutive item pairs (bigrams) to the total number of pairs:
\begin{align}
        \text{Distinct-2} = \frac{\text{\# unique bigrams}}{\text{\# total bigrams}},
\end{align}
A higher value suggests richer within-list variation.
In practice, online reranking systems generally employ multiple generators~\cite{yang2025comprehensive} to provide diverse candidate lists. Specifically, we adopt the generative model and the Ensemble Sort (ES) strategy~\cite{Sener2018MultiTaskLAensemblesort} as distinct parallel generation paths.
These generated sequences compete for exposure eligibility that are ultimately exposed to users.
As shown in~\autoref{tab:diversity_results}, our \methodname achieves the highest diversity score, item coverage, and the lowest repetition rate, demonstrating its strong capability to produce more personalized sequences with minimal item redundancy. Besides, our highest distinct-2 indicates more diverse and less repetitive local patterns.
\begin{table}[ht]
\setlength{\tabcolsep}{3pt}
\centering
\caption{Comparison of cross-list diversity among different generators for the Kuaishou dataset.}
\begin{tabular}{l|cccc}
\toprule
Method & \makecell[c]{Diversity\\Score} 
& \makecell[c]{Repetition\\Rate} 
 &\makecell[c]{Item\\Coverage} & \makecell[c]{Distinct-2} \\
\midrule
Ensemble Sort & 0.59 & 55.68\%  &26.67\% & 52.86\% \\
NAR4Rec & 0.74 & 33.43\%  &64.33\% & 11.80\% \\
\textbf{\methodname (Ours)} & \textbf{0.85} & \textbf{23.25\%}  &\textbf{72.57\%} & \textbf{65.52\%} \\
\bottomrule
\end{tabular}
\label{tab:diversity_results}
\end{table}

This observation also provides empirical evidence for the existence of the \textit{likelihood trap}. Specifically, NAR4Rec corresponds to the generative reranking model before introducing our framework, whereas \methodname can be viewed as its enhanced version after alleviating the likelihood trap. Compared with NAR4Rec, \methodname consistently improves item coverage and Distinct-2 while further reducing repetition, suggesting that the original generative model tends to repeat frequent local item patterns, while our method encourages more diverse and consistent list generation.

\noindent
\textbf{Effect of Graph Size Factor  }
The decoder structures hidden states into a directed acyclic graph, where the vertex count $g$ scales with target length $m$ by a graph size factor $\lambda$ (i.e., $g = \lambda \times m$).
Intuitively, a larger $\lambda$ provides more intermediate vertices and thus expands the space of feasible transition paths, allowing the decoder to explore more diverse candidate list structures. As shown in~\autoref{fig:graphsizefactor}, Recall@6 generally improves as $\lambda$ increases and reaches the best performance at $\lambda = 6$. 
Further increasing $\lambda$, however, brings no extra benefit and may introduce redundant vertices, leading to slight performance degradation.

\begin{figure}[ht]
    \centering
    \begin{tikzpicture}
        \begin{axis}[
            width=8cm, height=5.5cm,
            axis lines=left, 
            axis line style={thick, black}, 
            grid=both,
            grid style={dashed, gray!30},
            xlabel={Graph Size Factor ($\lambda$)},
            ylabel={Recall@6 (\%)},
            xmin=1, xmax=13.5,
            ymin=64, ymax=75.5,
            xtick={0, 2, 4, 6, 8, 10, 12},
            ytick={66, 68, 70, 72, 74},
            label style={font=\large}, 
            tick label style={font=\large},
            major tick length=0pt,
        ]

        \addplot[
            color=scienceblue,
            mark=*,
            mark size=1.5pt,
            line width=1.2pt, 
        ] coordinates {
            (2, 66)
            (4, 73)
            (6, 73.7)
            (8, 73)
            (10, 72.7)
            (12, 72.5)
        };


        \end{axis}
    \end{tikzpicture}
    \caption{Impact of the graph size factor ($\lambda$) on Recall@6 for the Kuaishou dataset. }
            \label{fig:graphsizefactor}
\end{figure}

We further evaluate the online inference latency under different graph size factors. As shown in~\autoref{tab:graph_size_online_latency}, increasing $\lambda$ only introduces negligible latency changes in the online serving systems. 
This is because both the transition matrix and the item prediction matrix are computed in a highly parallel manner. The only sequential component is a for-loop over the output length $m$; however, $m$ is typically small in practical scenarios, usually less than ten~\cite{ren2024non}. 
Therefore, increasing the graph size does not introduce substantial online computation overhead. 
Notably, the slight non-monotonic latency variations are mainly due to normal fluctuations in the online serving environment, such as batching, scheduling, and system load.
Considering that $\lambda = 4$ already achieves strong Recall@6 while maintaining a compact graph structure and stable online latency, we adopt $\lambda = 4$ as the default setting.

\begin{table}[ht]
\setlength{\tabcolsep}{2pt}
\centering
\caption{Online inference latency under different graph size factors.} 
\begin{tabular}{l|ccccccc}
\toprule
{Graph Size Factor} ${(\lambda)}$
        & {$1$} & {$2$} & {$4$} & {$6$} & {$8$} & {$10$} & {$12$} \\
\midrule
{Online Latency (ms)}
        & 20.60 & 20.68 & 20.27 & 20.67 & 20.58 & 21.25 & 21.27 \\
{$\Delta$ Latency (ms)}
        & -- & +0.08 & -0.33 & +0.07 & -0.02 & +0.65 & +0.67 \\
\bottomrule
\end{tabular}
\label{tab:graph_size_online_latency}
\end{table}


\subsection{Ablation Study} \label{sec:ablation}


We perform extensive ablation studies to isolate the effectiveness of the core designs in \methodname, including the Consistent Differentiable Training method and the proposed decoder architecture. By removing or replacing key components, we analyze their individual contributions to both offline and online recommendation performance. The results further validate the necessity of each design choice and provide a fine-grained understanding of where the improvements come from.

\noindent
\textbf{Ablation on Consistent Differentiable Training method  }
To verify the necessity of the consistent alignment objective, we compare the performance of the model trained solely with the generative loss ($\mathcal{L}_{gen}$) against the complete model that uses the joint objective $(\mathcal{L}_{total})$. The results are summarized in~\autoref{tab:training_methods}. As observed, optimizing solely \(\mathcal{L}_{gen}\) boosts \textit{Views} (+0.979\%) but degrades \textit{Watch Time} (-0.150\%), suggesting a tendency to recommend items that attract views but fail to sustain engagement (i.e., ``clickbait'').
In contrast, our combined objective \(\mathcal{L}_{total}\) aligns generation with the evaluator's utility, securing a robust increase in \textit{Views} (+0.780\%) while reversing the decline in \textit{Watch Time} (+0.109\%) to achieve a superior balance between view quantity and user retention.
\begin{table}[ht]
\centering
\caption{
Ablation study on Consistent Differentiable Training. 
Values represent relative improvements over the online baseline NAR4Rec.
\textcolor{ForestGreen}{Green} indicates improvement (positive), while \textcolor{red}{Red} indicates degradation (negative).
}
\begin{tabular}{l|cc}
\toprule
  Method& Views & Watch Time \\
\midrule
 \textit{Only} \(\mathcal{L}_{gen}\) & \textcolor{ForestGreen}{+0.979\%} & \textcolor{red}{-0.150\%} \\
 \textbf{\(\mathcal{L}_{total}\) (Ours)} & \textcolor{ForestGreen}{+0.780\%} & \textcolor{ForestGreen}{+0.109\%} \\
\bottomrule
\end{tabular}
\label{tab:training_methods}
\end{table}

\nop{
\begin{table}[ht]
\centering
\caption{
Ablation study on training objectives. 
All values represent the relative improvements over the online baseline. 
\textcolor{red}{Red} and \textcolor{ForestGreen}{Green} values indicate significantly positive and negative feedback, respectively.
(A relative change of 0.2\% is typically considered significant.)}
\begin{tabular}{c|cc}
\toprule
  & Views & Watch Time \\
\midrule
 \textit{Only} \(\mathcal{L}_{gen}\) & \textcolor{red}{+0.853\%*} & \textcolor{ForestGreen}{-0.801\%*} \\
 \textit{Only} \(\mathcal{L}_{con}\)& +0.105\% & {+0.185\%} \\
 \textbf{\(\mathcal{L}_{total}\) (Ours)} & \textcolor{red}{+0.780\%*} & -0.021\%\\
\bottomrule
\end{tabular}
\label{tab:training_methods}
\end{table}
}
\nop{
\begin{table}[ht]
\centering
\caption{
Ablation study on Consistent Differentiable Training. 
All values represent relative improvements over the online baseline.
\textcolor{red}{Red} and \textcolor{ForestGreen}{Green} values indicate significantly positive and negative feedback, respectively.
}
\begin{tabular}{c|cc}
\toprule
  & Views & Watch Time \\
\midrule
 \textit{Only} \(\mathcal{L}_{gen}\) & \textcolor{red}{+0.979\%}& \textcolor{ForestGreen}
{-0.150\%}\\
 \textit{Only} \(\mathcal{L}_{con}\)& \textcolor{red}{+0.305\%*}& \textcolor{red}{+0.598\%*}\\
 \textbf{\(\mathcal{L}_{total}\) (Ours)} & \textcolor{red}{+0.780\%}& \textcolor{red}{+0.109\%}\\
\bottomrule
\end{tabular}
\label{tab:training_methods}
\end{table}
}

\noindent
\textbf{Ablation on Decoder Architecture } 
Our \methodname integrates a novel graph-structured decoder to expand the decoding exploration space.
To verify its superiority, we compare our method against the vanilla decoder in~\cite{ren2024non} under the same consistent differentiable training framework.
\autoref{tab:decoderablation} reveals a significant performance drop with the vanilla decoder, attributed to its restriction to linear chains. Conversely, our graph-structured design expands the decoding space and captures richer dependencies, thereby producing higher-quality sequences and superior engagement metrics.

\begin{table}[ht]
    \centering
    \setlength{\tabcolsep}{3pt} 
    \caption{Ablation study on decoder architectures comparing both online (\textit{Views} and \textit{Watch Time}) and offline (Recall) performance for the Kuaishou dataset. We report relative values over the baseline NAR4Rec.
    }
    \begin{tabular}{l|cc|cc}
    \toprule
    Method& \multicolumn{2}{c}{\textbf{Online}} & \multicolumn{2}{c}{\textbf{Offline}} \\
    \cmidrule(lr){2-3} \cmidrule(lr){4-5}
     & Views & Watch Time & Recall@6 & Recall@10 \\
    \midrule
    Vanilla Decoder & +0.310\% & -0.075\% & 68.41\% & 75.20\% \\
    \textbf{Ours} & \textbf{+0.780\%} & \textbf{+0.109\%} & \textbf{72.84\%} & \textbf{81.83\%} \\
    \bottomrule
    \end{tabular}
    \label{tab:decoderablation}
\end{table}



\section{Conclusion}
In this work, we introduce \methodname, a novel consistent graph-structured generative reranking framework. To the best of our knowledge, we are the first to identify the existence of the \textit{likelihood trap} in generative recommender systems and provide effective solutions. 
Specifically, we first introduce a novel Graph-structured Model, which enables the generation of more diverse sequences by substantially expanding the decoding exploration space. Building on this, we explicitly model item dependencies via a learnable graph transition module to improve prediction accuracy.
Furthermore, we propose a novel Consistent Differentiable Training method, which incorporates the evaluator into the generator's training to guide it toward alignment with user preferences.
We conduct extensive offline and online experiments to demonstrate the effectiveness of \methodname, and the results show that it significantly improves both the quality and diversity of generated sequences.

\bibliographystyle{ACM-Reference-Format}
\balance
\bibliography{sample-base}










\end{document}